\documentclass[useAMS]{mn2e}
\usepackage{psfig}

\usepackage{times}
\def\spose#1{\hbox to 0pt{#1\hss}}
\newcommand\lsim{\mathrel{\spose{\lower 3pt\hbox{$\mathchar"218$}}
     \raise 2.0pt\hbox{$\mathchar"13C$}}}
\newcommand\gsim{\mathrel{\spose{\lower 3pt\hbox{$\mathchar"218$}}
     \raise 2.0pt\hbox{$\mathchar"13E$}}}
\def\ltsima{$\; \buildrel < \over \sim \;$}
\def\lsim{\lower.5ex\hbox{\ltsima}}
\def\gtsima{$\; \buildrel > \over \sim \;$}
\def\gsim{\lower.5ex\hbox{\gtsima}}

\def\sch{Schwarzschild}

\title[Compton rockets and relativistic jets]
{Compton rockets and the minimum power of relativistic jets}

\author[Ghisellini \& Tavecchio] 
{G. Ghisellini\thanks{E--mail:
gabriele.ghisellini@brera.inaf.it} 
and F. Tavecchio
\\
INAF -- Osservatorio Astronomico di Brera, via E. Bianchi 46, I--23807 Merate, Italy \\
}

\begin{document}


\pagerange{\pageref{firstpage}--\pageref{lastpage}} \pubyear{2007}

\maketitle

\label{firstpage}

\begin{abstract}
The power of a relativistic jet depends on the number of leptons and protons 
carried by the jet itself.
We have reasons to believe that powerful $\gamma$--ray flat spectrum radio sources
emit most of their radiation where radiative cooling is severe.
This helps to find the minimum number of emitting leptons needed to
explain the radiation we see. 
The number of protons is more uncertain.
If there is one proton per electron, they dominate the
jet power, but they could be unimportant 
if the emission is due to electron--positron pairs.
In this case the total jet power could be much smaller.
However, if the $\gamma$--ray flux is due to 
inverse Compton scattering with seed photons
produced outside the jet, the radiation is anisotropic 
also in the comoving frame, making the jet to recoil.
This Compton rocket effect is strong for light, electron--positron jets, 
and negligible for heavy, proton dominated jets.
No significant deceleration, required by
fast superluminal motion, requires a minimum number
of protons per lepton, and thus a minimum jet power.
We apply these ideas to the blazar 3C 454.3, to 
establish a robust lower limit to its total jet power: if the
viewing angle $\theta_{\rm v}\sim 1/\Gamma$ the 
jet power is larger than the accretion luminosity $L_{\rm d}$ for any 
bulk Lorentz factor $\Gamma$.
For $\theta_{\rm v}=0^\circ$, instead, the minimum jet power can
be smaller than $L_{\rm d}$ for $\Gamma<25$.
No more than $\sim$10 pairs per proton are allowed.
\end{abstract}
\begin{keywords}
galaxies: active--galaxies: jets--galaxies: individual: 3C454.3 ---
radiation mechanisms: non--thermal
\end{keywords}

\section{Introduction}

The high quality data of the the Large Area Telescope (LAT) onboard the {\it Fermi} 
satellite, together with the simultaneous observations performed by the 
{\it Swift} satellite in the optical--UV and X--ray bands and by 
ground based telescopes allowed a new era in the study of blazar jets.
Detailed modelling of these sources allows to
estimate the physical parameters of the jet emitting region, such
as its magnetic field, particle density, size and bulk Lorentz factor.
Therefore we can estimate the power that the jet
carries in the form of particles and fields, and  compare it
with the accretion luminosity, at least in Flat Spectrum Radio Quasars
(FSRQs) where the disk component is visible.
In our previous studies of $\gamma$--ray loud FSRQs 
(Ghisellini et al. 2010a, 2010b) we found that the jet power can be even 
larger than the accretion luminosity, and it correlates with it (also when
accounting for the common redshift dependence).

When estimating the jet power in this way there
are two crucial uncertainties: i) the total number of leptons,
that depends on 
low energy end of the particle distribution (as those are most numerous) yet difficult 
to observe because of synchrotron self--absorption,
and ii) the number of protons per lepton.
For the first concern, evidence is accumulating that in FSRQs the radiative
cooling is severe, so that leptons of almost all energies do cool in one
light crossing time, and the presence of low energy particles is often required
to reproduce the observed X--ray spectrum, interpreted as inverse Compton radiation
with photons originating externally to the jet (i.e. External Compton, EC hereafter).

The second concern (how many pairs per proton) has been discussed,
among others, by Ghisellini et al. (1992);
Celotti \& Fabian (1993);
Sikora \& Madejski (2000); Celotti \& Ghisellini (2008); Ghisellini et al. (2010a). 
If pairs are created in the $\gamma$--ray emission
region we should see a clear break in the spectrum,
and the absorbed luminosity should be reprocessed at lower energies, especially
in the X--ray band, where instead the spectral energy distribution (SED) 
of FSRQs has a minimum.
If the pairs are created very close to the black hole, there is a maximum 
number of them surviving annihilation, corresponding to a local pair 
scattering optical depth $\tau_\pm\sim 1$ (Ghisellini et al. 1992).
When arriving to the parsec VLBI scale, the corresponding pair density is
less than the lepton density required to produce the synchrotron flux we see.
On the other hand, the $\gamma$--ray emitting zone is much smaller and closer to the
black hole than the VLBI zone, and the number density of the surviving pairs
might be enough to account for the radiation produced in this region.
We have found in our earlier works (Celotti \& Ghisellini 2008; Ghisellini et al. 2010a,b)
that the power spent by the jet to produce its radiation is often greater than the power
in Poynting flux and bulk kinetic energy of the emitting leptons, requiring an
additional form of jet power, i.e. protons.
The simplest hypothesis of one proton per electron leads to 
jet powers systematically larger than the accretion luminosity.
So it is crucial to evaluate how many protons per emitting lepton there
are in the jet.
To this end we introduce here a new argument, that can be applied when
most of the radiation is produced through Compton scattering with
external radiation (EC). 
In this case, the emission pattern is anisotropic in the comoving frame
of the emitting region, that must recoil.
This is the ``hot version" of the Compton drag effect (the emitting
particles are relativistic in the comoving frame) and is called 
``Compton rocket" (CR hereafter) effect.
First studied in the `80s, (O'Dell 1981) as a way to accelerate jets
using the accretion disk photons as seeds, it has been used 
as a tool to limit the jet bulk Lorentz factor $\Gamma$ assuming, as seed photons,
those re--isotropized by the broad line region (BLR) or
by a relatively distant torus (e.g. Sikora et al. 1996).
More recently, it has been used as a decelerating agent for very fast jet ``spines" 
moving inside slower jet ``layers" (Ghisellini et al. 2005),
or for large scale jets interacting with the cosmic microwave background 
(Tavecchio et al. 2006).

We use the CR effect to limit the number of pairs, assuming
that the jet is moving with a given $\Gamma$, and requiring it does not
significantly decelerate by the CR effect, in order to be consistent
with observations of fast superluminal motion at the VLBI scales.
Light jets (i.e. pair dominated) can be decelerated more effectively
than heavier jets (i.e. with an important proton component).
Therefore requiring no significant deceleration fixes the minimum number of protons 
per lepton, one of the most important number to find out a limit on the
total jet power.
Furthermore, 
if the (energetically dominant) $\gamma$--ray flux is EC emission, 
we can in a rather straightforward way evaluate the jet powers
in its different forms: magnetic, leptonic, radiative and protonic,
and how these different jet powers change by changing $\Gamma$
(similarly to what done in Ghisellini \& Celotti 2001).
Doing this, one finds a {\it minimum} power, approximately where
the Poynting flux equals the other dominant form of power (i.e.
bulk motion of leptons, or protons, or radiative), and a corresponding
$\Gamma$.

We apply these arguments to 3C 454.3, one of the best studied blazars,
used as a test case.
It is a FSRQ at $z=0.859$ (Jackson \& Browne 1991), superluminal
(Jorstad et al. 2005;  Lister et al. 2009) with components moving
with $\beta_{\rm app}$ from a few to more than 20, 
resulting in estimated bulk Lorentz factors from 10 to 25.
It is one of the brightest and most variable FSRQs.
In  April--May 2005 it underwent outburst, dramatic in optical 
(Villata et al. 2006) and visible also at X--ray energies  
(Pian et al. 2006; Giommi et al. 2006). 
The {\it AGILE} satellite detected 3C 454.3 as one of the brightest
and variable sources in the $\gamma$--ray band (Vercellone et al. 2007, 2010).
After the launch of the {\it Fermi} satellite 
3C 454.3 was seen to flare several times (Tosti et al. 2008; Abdo et al. 2009), 
with a climax in December 2009 (Bonnoli et al. 2010).


\section{The Compton rocket effect and the pair content of the jet}

If the main emission process of blazars is synchrotron and self--Compton
radiation, then the emitted luminosity is isotropic in the comoving frame.
This means that the jet loses mass, but not velocity.
The lost mass is at the expenses of the ``relativistic" (i.e. $\gamma m_{\rm e} c^2$)
mass of the emitting leptons. 
Instead, if powerful blazars produce most of their emission by
scattering {\it external} radiation, the produced radiation
is anisotropic even in the comoving frame, 
and the jet must decelerate (i.e. it recoils in the comoving frame).
The amount of this deceleration depends on the produced EC luminosity
and by the inertia of the jet, i.e. if the jet is ``heavy" or ``light".
Therefore this ``Compton rocket" (CR)
effect (i.e. the ``hot" version of Compton drag, because the leptons are
relativistic) can give some limit on the minimum number of protons
present in the jet, requiring that is does not decelerate significantly.
We here present a simple derivation of the relevant formulae, that
agree with the more detailed and complex derivation of Sikora et al. (1996).

Let us remain in the observer frame.
There we measure an external and isotropic radiation energy density $U_{\rm ext}$.
The total Lorentz factor ($\tilde\gamma$) of the electrons is the superposition of the 
bulk ($\Gamma$) and random ($\gamma$) Lorentz factors ($\beta_{\rm bulk}$ and $\beta$ are
the corresponding velocities). 
Assume that the bulk motions occurs along the x--axis, and that the random velocity 
forms an angle $\theta^\prime$ with respect to that axis, in the comoving frame.
We have (e.g.  Rybicki \& Lightman 1979):
\begin{equation}
\beta_x ={ \beta^\prime\cos\theta^\prime +\beta_{\rm bulk} 
\over 1+\beta_{\rm bulk}\beta^\prime\cos\theta^\prime};
\quad
\beta_y ={ \beta^\prime\sin\theta^\prime  \over 
\Gamma (1+\beta_{\rm bulk}\beta^\prime\cos\theta^\prime )}
\end{equation}
The total $\tilde\gamma^2$ is
\begin{equation}
\tilde\gamma^2 \,  = \, \left[ 1-\beta_x^2-\beta_y^2\right]^{-1} \, =\,
(1+\beta_{\rm bulk}\beta^\prime\cos\theta^\prime )^2 \gamma^2\Gamma^2 
\end{equation}
If the particle distribution is isotropic in the comoving frame,
the average over angles gives
%
\begin{equation}
\langle \tilde\gamma^2\rangle 
= { \int 2\pi \sin\theta^\prime \tilde\gamma^2(\theta^\prime) d\theta^\prime \over 4\pi }
= \left[ 1+{(\beta_{\rm bulk}\beta^\prime)^2 \over 3}\right] \gamma^2\Gamma^2
\label{g2}
\end{equation}
which gives the factor $(4/3)$ for ultra--relativistic speeds.

Now assume that a portion of the jet carries a total number $N_{\rm p}$ of protons
and $N_{\rm e}$ leptons (including pairs). 
The ``cooling time" of the jet (i.e. the time for halving $\Gamma$) is
\begin{eqnarray}
t_{\rm cool} &= & {E\over \dot E} = \Gamma \,\, 
{ N_{\rm p} m_{\rm p} c^2 +N_{\rm e} \langle\gamma\rangle m_{\rm e} c^2 \over 
(4/3) \sigma_{\rm T} c N_{\rm e} U_{\rm ext} \langle \tilde \gamma^2\rangle  }   \nonumber \\
 & = &
{9\over 16} \, { (N_{\rm p}/ N_{\rm e}) m_{\rm p} c^2 + \langle \gamma \rangle m_{\rm e} c^2 
\over \sigma_{\rm T} c U_{\rm ext} \langle \gamma^2\rangle \Gamma  }
\end{eqnarray}
In the cooling time $t_{\rm cool}$, the jet travels a distance $R_{\rm cool}=\beta c t_{\rm cool}$.
The corresponding interval of time as measured by the observer
is Doppler contracted by the factor $(1-\beta\cos\theta_{\rm v})\equiv 1/(\Gamma\delta)$
($\theta_{\rm v}$ is the viewing angle and $\delta$ the beaming factor).
The time $t_{\rm cool}(1-\beta\cos\theta_{\rm v})$ 
has to be compared with the timescale for which the particles
are indeed described by an energy distribution with the value of $\langle \gamma^2\rangle$
used, provided that, during this time, the radiation density remains $U_{\rm ext}$.
This timescale is approximately the variability timescale.
The CR effect is unimportant if
\begin{eqnarray}
t_{\rm cool}
& > &
{t_{\rm var}\over (1+z) (1-\beta\cos\theta_{\rm v})} = {t_{\rm var}\Gamma \delta \over (1+z) }\to  \nonumber \\
\Gamma^2  \delta  &<&
{9  (1+z) \over 16} \,\, { (N_{\rm p}/ N_{\rm e}) m_{\rm p} c^2 + \langle \gamma \rangle m_{\rm e} c^2 
 \over \sigma_{\rm T} c  U_{\rm ext} \langle \gamma^2\rangle   t_{\rm var}}
\label{cd}
\end{eqnarray}
This limit becomes very severe if the jet is dominated by 
hot pairs and if $U_{\rm ext}=U_{\rm BLR}$, the radiation energy density
is dominated by radiation from the broad line region.
In this case jets with $\Gamma\gsim 10$
are bound to decelerate.
They do not decelerate if they contain a proton component that increases their inertia.
We can rewrite Eq. \ref{cd} to find the minimum ratio $N_{\rm p}/N_{\rm e}$
compatible with halving $\Gamma$ in $t_{\rm var}$:
\begin{equation}
{ N_{\rm p}\over N_{\rm e} } > \max \left[ 0, \,  
\left( {16 \over 9} {\Gamma^2 \delta t_{\rm var} \over 1+z} { \sigma_{\rm T} c  U_{\rm ext}
\langle\gamma^2\rangle\over m_{\rm p} c^2}  - 
{ \langle\gamma\rangle m_{\rm e} \over m_{\rm p} } \right) \right]
\label{ratiop}
\end{equation}
Since $N_{\rm e}= N_\pm + N_{\rm p}$ (pairs plus electrons associated with protons),
we have $N_\pm/N_{\rm p} = (N_{\rm e}/N_{\rm p})-1$.

\section{The power of the jet}

The most robust estimate on the jet power 
is the power $P_{\rm r}$ spent to produce the radiation as
measured with a ``$4\pi$" detector surrounding the source
in the observer frame.
If the luminosity $L^\prime$ is isotropic in the comoving frame  
we would derive $P_{\rm r} = L^\prime/(4\pi)\int \delta^4 d\Omega =  (4/3)\Gamma^2 L^\prime$.
But if the EC process is important the emission is not isotropic in the rest frame,
and the observed flux, instead of being boosted by $\delta^4$, follows 
a pattern given by $\delta^4 (\delta/\Gamma)^2$ (see Dermer 1995 and 
Georganopoulos et al. 2001). 
Setting $\langle L^\prime \rangle$ the angle averaged luminosity in the comoving frame,
we then have
\begin{equation}
P_{\rm r} = {\langle L^\prime\rangle  \over 4 \pi}\,\int_{4\pi} {\delta^6(\theta) \over \Gamma^2} d\Omega \sim 
{16\over 5} \Gamma^2 \langle L^\prime \rangle 
\approx
{16 \, \Gamma^4 L_{\rm obs} \over 5 \, \delta^6(\theta_{\rm v}) } 
\label{pr}
\end{equation}
%

The power in bulk motion of leptons, protons, and magnetic fields
are calculated as:
\begin{eqnarray}
P_{\rm e} &=&  \pi  R_{\rm blob}^2 \Gamma^2 \beta c \, 
m_{\rm e} c^2 \int^{\gamma_{\rm max}}_{\gamma_{\rm cool}} N(\gamma) \gamma d\gamma 
\nonumber \\
&=& \pi R_{\rm blob}^2 \Gamma^2 \beta c  \, n_{\rm e} \langle \gamma \rangle   m_{\rm e} c^2
\nonumber \\
P_{\rm p} &=&  \pi R_{\rm blob}^2 \Gamma^2 \beta c \, n_{\rm p} m_{\rm p} c^2 
\nonumber \\
P_{\rm B} &=&  \pi R_{\rm blob}^2 \Gamma^2 \beta c \, U_{\rm B}
\label{powers}
\end{eqnarray}
where $R_{\rm blob}$ is the size of the emitting source.
We are assuming that protons are cold and that $n_{\rm e}$ 
is the total number density of leptons, including pairs (if present),
so that $n_{\rm e} =n_\pm+n_{\rm p}$.
We also assume that {\it all} leptons are relativistic 
and are described by the energy distribution $N(\gamma)$.
Neglecting cold leptons minimizes the power requirement.
The total jet power is $P_{\rm jet} = P_{\rm r}+P_{\rm e}+P_{\rm p}+P_{\rm B}$.

As long as the scattering is in the Thomson regime
the observed luminosity in the EC component of the SED is:
\begin{equation}
L_{\rm EC}^{\rm obs} 
\sim 
{16\pi R_{\rm blob}^3 \over 9} \sigma_{\rm T} c n_{\rm e}\langle\gamma^2\rangle 
U^\prime_{\rm ext} {\delta^2 \over \Gamma^2}\, \delta^4 
\end{equation}
We can then find the number density $n_{\rm e}$ of the emitting leptons:
\begin{equation}
n_{\rm e} ={9 L_{\rm EC} \over 
16\pi R_{\rm blob}^3 \sigma_{\rm T} c \, \langle \gamma^2\rangle  U_{\rm ext}\delta^6}
\end{equation}
The averages $\langle\gamma\rangle$ and $\langle\gamma^2\rangle$
are calculated assuming the emitting particle distribution is
a broken power law,
extending from $\gamma_{\rm cool}$ to $\gamma_{\rm peak}$ with slope
$N(\gamma)\propto \gamma^{-2}$, as appropriate for radiative cooling,
and breaking above $\gamma_{\rm peak}$, where we assume $N(\gamma)\propto \gamma^{-p}$
up to $\gamma_{\rm max}$.
The slope $p$ is related to the observed energy spectral index $\alpha$ above the synchrotron and
the EC peaks as $p=2\alpha+1$.
The values of $\gamma_{\rm cool}$ and $\gamma_{\rm peak}$ will depend on $\Gamma$ and $\delta$
(see Eq. \ref{gcool} and Eq. \ref{gpeak}).
In general, both $\langle \gamma\rangle$ and $\langle\gamma^2\rangle$ decrease by increasing
$\Gamma$ and $\delta$: this is because the external photon field is seen more boosted
in the comoving frame, inducing a stronger Compton cooling (and so $\gamma_{\rm cool}$ decreases);
at the same time a smaller $\gamma_{\rm peak}$ is required to produce the high energy peak.

The magnetic energy density $U_{\rm B}$ can be derived
in terms of the ``Compton dominance'', namely the $\gamma$--ray to synchrotron luminosity
ratio $L_\gamma/L_{\rm syn}$:
\begin{equation}
{L_\gamma \over L_{\rm syn} }   =  
{U^\prime_{\rm ext} (\delta/\Gamma)^2 \over U_{\rm B} } 
\to  U_{\rm B} =   \delta^2 U_{\rm ext} \,   {L_{\rm syn}\over L_\gamma } 
\label{B}
\end{equation}

\section{The minimum jet power}

We now show how the different forms of jet power changes
by changing the bulk Lorentz factor.
These estimates depends on the following parameters:
$L_\gamma\sim L_{\rm EC}$;
$L_\gamma / L_{\rm syn}$; 
$\nu_{\rm c}$ (the $\nu F_\nu$ peak frequency of the $\gamma$--ray spectrum); 
$\theta_{\rm v}$ (the viewing angle);
$t_{\rm var}=(1+z)R_{\rm blob}/(c\delta)$; 
$\alpha_\gamma$ (the energy spectral index of the $\gamma$--ray spectrum above the peak) and 
$z$.
Of these 7 parameters, all but one (the viewing angle) are observables.
Note that the external radiation energy density is not a free parameter
if the typical radius of the BLR (or the reprocessing torus) depends
on the disk luminosity as $R_{\rm ext}\propto L_{\rm d}^{1/2}$. 
In this case $U_{\rm ext}\propto L_{\rm d}/R^2_{\rm ext}$ is constant.

The viewing angle $\theta_{\rm v}$ is in general unknown.
We can assume $\theta_{\rm v}=1/\Gamma$, in order to always have $\Gamma=\delta$.
Alternatively, we can assume $\theta_{\rm v}=0$, i.e. $\delta=2\Gamma$.
Maximizing the Doppler boosting, this choice will minimize the derived powers
but the probability to observe any source with $\theta_{\rm v}=0$ is vanishingly small.
There might be an exception: consider the case for which
the velocity vectors of the emitting flow are not parallel,
but somewhat radial within the jet aperture angle $\theta_{\rm jet}$. 
Thus if $\theta_{\rm v}$ is finite, but smaller than $\theta_{\rm jet}$, 
there is a {\it portion} of the jet exactly pointing at us.
On the other hand, the corresponding emitting volume of this portion is small,
and the flux we see is mostly contributed for by those parts of the jet
moving with $\theta_{\rm v} \sim 1/\Gamma$, because they have a larger volume.

We are led to conclude that the case $\theta_{\rm v} = 1/\Gamma$ is favored.
It is also the angle for which the superluminal motion is maximized, but this
implies that the jet does not change direction between the $\gamma$--ray emitting
region and the VLBI scale, that is not guaranteed.

We now rewrite the different forms of jet power in order to make more transparent
their dependences on the bulk Lorentz factor and the parameters listed above.
\begin{eqnarray}
P_{\rm B} &=&  {\pi c^3 t^2_{\rm var}  \over  (1+z)^2} \,
{L_{\rm syn} \over L_\gamma} \, U_{\rm ext}\beta \Gamma^4\delta^2
\nonumber \\
P_{\rm e} &=& {9  \over 16} {\langle \gamma \rangle \over \langle \gamma^2 \rangle }
{ m_{\rm e} c^2 \over \sigma_{\rm T} c U_{\rm ext} } { (1+z) L_\gamma \over t_{\rm var} }\, \delta^{-5}
\nonumber \\
P_{\rm p} &=&  P_{\rm e} \, {m_{\rm p} \over  \langle \gamma \rangle m_{\rm e} }; 
\qquad \quad \,\,\, {n_{\rm p}=n_{\rm e}}
\nonumber \\
P_{\rm p} &\sim &  L_\gamma \,\, \Gamma^4 \delta^{-6} \sim P_{\rm r};
\quad n_{\rm p}\,\,  {\rm from \,\, Eq.\,\, \ref{ratiop}}
\label{powers}
\end{eqnarray}
The last approximate equality assumes that the second term in round brackets 
in Eq. \ref{ratiop} is negligible with respect to the first.
Note that, as expected, the minimum $P_{\rm p}$ limited by the
CR effect is of the order of the power spent in radiation
(given in Eq. \ref{pr}).
Be aware that $\langle \gamma \rangle$ and $\langle \gamma^2 \rangle$ do depend
on $\Gamma$ (see Eq. \ref{gcool} and Eq. \ref{gpeak} below).

\begin{figure}
\hskip -1cm
\psfig{file=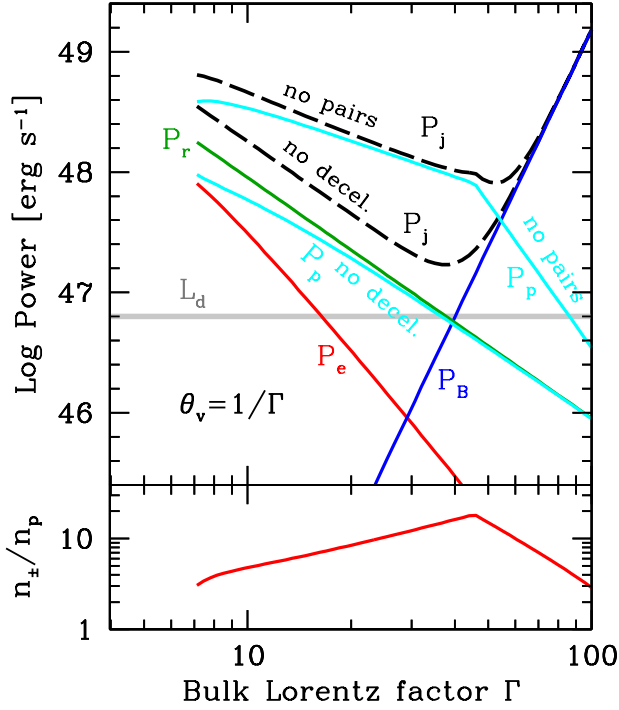,height=10.5cm,width=10.5cm}
\vskip -0.5cm
\caption{Different form of jet power as a function of 
the bulk Lorentz factor $\Gamma$, as labelled.
For this particular example, we have assumed 
$L_\gamma=9\times 10^{49}$ erg s$^{-1}$, $\nu_{\rm c} = 10^{22}$ Hz,
$t_{\rm var}=6$ hours and $L_\gamma/L_{\rm syn}=20$, appropriate for the blazar 3C 454.3,
at $z=0.859$.
We have further assumed that $\delta=\Gamma$, implying 
$\theta_{\rm v}=1/\Gamma$.
The horizontal grey line indicates the accretion disk luminosity.
The long dashed lines correspond to $P_{\rm jet}$ assuming
one proton per emitting electrons (i.e. no pairs) or 
instead assuming the minimum number of protons per electron
consistent with no strong jet deceleration for the Compton rocket effect. 
We can see that $P_{\rm jet}>L_{\rm d}$ for all $\Gamma$.
The bottom panel shows the maximum pair to proton ratio allowed by the CR effect.
}
\label{lmin}
\end{figure}

\begin{figure}
\hskip -0.7cm
\psfig{file=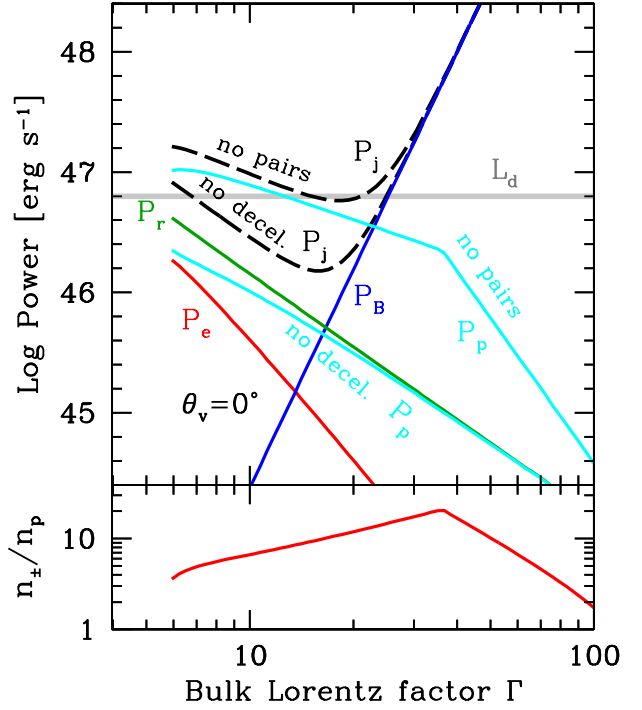,height=10.5cm,width=10.5cm}
\vskip -0.5cm
\caption{Same as Fig. \ref{lmin}, but assuming
$\theta_{\rm v}=0$. Note the different range of the y axis.
In the bottom panel, the function $n_\pm/n_{\rm p}$ is similar,
but not identical to the one shown in Fig. \ref{lmin}. 
}
\label{lmin0deg}
\end{figure}

\subsection{Application to the blazar 3C 454.3}

In order to find a lower limit to the jet power of 3C 454.3 we make
the following assumptions:
\begin{itemize}
\item
In Bonnoli et al. (2010) we have shown that during the big flare of Nov--Dec 2009 the $\gamma$--ray,
X--ray and optical fluxes of 3C 454.3 were correlated with one another,
with the $\gamma$--ray flux varying more than linearly with the flux in the other two bands.
We take this as a very robust indication that most of the non--thermal flux
received from 3C 454.3 above the far IR band is produced in the same region of the jet.

\item
The size of the emitting region $R_{\rm blob}$ is assumed to be associated 
with the minimum variability timescale $t_{\rm var}$ of the source.
In the $\gamma$--rays, Tavecchio et al. (2010); Foschini et al. (2010) and 
Ackermann et al. (2010) found significant variations in 3--6 hours.
Therefore 
\begin{equation}
R_{\rm blob}= c t_{\rm var} {\delta\over 1+z }  
\sim 7\times 10^{15}\, \left( {t_{\rm var}\over 6\, {\rm hr} }\right) \,
\left( {\delta\over 20}\right) 
\quad {\rm cm}
\end{equation}

\item
We assume an accretion disk luminosity $L_{\rm d} \sim 6.7 \times 10^{46}$ erg s$^{-1}$,
based on direct detection of the Lyman--$\alpha$ line (Bonnoli et al. 2010)
and on the flattening of the optical--UV SED when the source is in low state.

\item
We assume that the BLR reprocesses 10\% of $L_{\rm d}$
and that the BLR size is given by
$R_{\rm BLR}= 10^{17} L^{1/2}_{\rm d, 45}$ cm.
This choice  (in rough agreement with Bentz et al. 2006 and Kaspi et al. 2007),
implies that the radiation energy density within the broad line region is constant:
\begin{equation}
U_{\rm BLR}= {0.1 L_{\rm d}\over 4\pi R_{\rm BLR}^2c} = 
{1\over 12 \pi} \quad {\rm erg\, cm^{-3}}
\end{equation}
The short $t_{\rm var}$ suggests that dissipation takes place within the BLR,
so we assume $U_{\rm ext}=U_{\rm BLR}$.

\item
After one light crossing time $R_{\rm blob}/c=t_{\rm var}\delta/(1+z)$ 
the cooling energy $\gamma_{\rm cool}$ is
\begin{equation}
\gamma_{\rm cool} \sim  { 3 (1+z) m_{\rm e}c^2 \over 
4\sigma_{\rm T} c  t_{\rm var}\, \delta \, 
[U^\prime_{\rm BLR}+U_B+U^\prime_{\rm syn}] }
\propto {1\over \Gamma^2\delta}
\label{gcool}  
\end{equation}
with the EC mechanism (with BLR photons as seeds) being the dominant cooling agent.

\item
The soft slope of the $\gamma$--ray spectrum and the hard slope of
the X--ray spectrum constrain the peak of the high energy component $h\nu_{\rm c}$
of the SED to lie close to 100 MeV.
For the EC process, the peak is made by electrons at $\gamma_{\rm peak}$
scattering the Ly--$\alpha$ seed photons with frequency $\nu_{Ly\alpha}$.
Then $\gamma_{\rm peak}$ is given by
\begin{equation}
\nu_{\rm c} = 2 \gamma_{\rm peak}^2 \nu_{Ly\alpha} {\Gamma\delta\over 1+z} 
\to
\gamma_{\rm peak} = \left[ {\nu_{\rm c} (1+z)
\over 2 \nu_{Ly\alpha} \,  \delta\, \Gamma }\right]^{1/2}
\label{gpeak}
\end{equation}
For all reasonable parameters appropriate for 3C 454.3, 
$\gamma_{\rm peak} > \gamma_{\rm cool}$, implying
that most of the energy injected in the form of relativistic leptons
is radiated away in one light crossing time.
\end{itemize}

The top panel of Fig. \ref{lmin} shows the different jet powers as a function of $\Gamma$ 
assuming that $\theta_{\rm v}=1/\Gamma$.
The two dashed lines are the total $P_{\rm jet}$ derived
assuming one proton per electron (i.e. no pairs) 
or instead the minimum number of protons given by Eq. \ref{ratiop}.
In both cases $P_{\rm jet}>L_{\rm d}$ for all values of $\Gamma$.
The minimum $P_{\rm jet}$ is set by the equipartition between 
$P_{\rm p}$ and $P_{\rm B}$.
This occurs at $\Gamma\sim 55$ for the ``no pairs" case, and $\Gamma\sim 40$
for the case of the minimum number of protons (``no decel." case).

The bottom panel of Fig. \ref{lmin} shows the maximum ratio $n_\pm/n_{\rm p}$
required not to decelerate significantly for the CR effect, as a function of $\Gamma$.
To understand the behavior of this curve consider Eq. \ref{ratiop}.
If we neglect the second term, we have 
$n_\pm/n_{\rm p}\propto (\Gamma^2\delta \langle\gamma^2\rangle)^{-1}$.
For illustration, consider a particle distribution extending
only between $\gamma_{\rm cool}$ and $\gamma_{\rm peak}$
with slope $N(\gamma)\propto \gamma^{-2}$.
In this case $\langle\gamma^2\rangle \sim \gamma_{\rm cool}\gamma_{\rm peak} 
\propto [\Gamma^2\delta (\Gamma\delta)^{1/2}]^{-1}$ as can be seen through Eq. \ref{gcool}
and Eq. \ref{gpeak}.
Therefore $n_\pm/n_{\rm p} \propto (\Gamma\delta)^{1/2} \propto \Gamma$.
This behavior ends when $\gamma_{\rm cool}$ becomes unity, i.e. for large
values of $\Gamma$. 
In this case $n_\pm/n_{\rm p} \propto \Gamma^{-1}(\Gamma\delta)^{-1/2}\sim \Gamma^{-2}$.
The maximum in $n_\pm/n_{\rm p}$ therefore occurs when $\gamma_{\rm cool}$ becomes unity.

The top panel of Fig. \ref{lmin0deg} shows the powers assuming $\theta_{\rm v}=0^\circ$.
In this case the minimum $P_{\rm jet}$ occurs for $\Gamma \sim 18$ (no pairs) or $\Gamma\sim 16$
(with $N_{\rm p}/N_{\rm e}$ given by Eq. \ref{ratiop}).
In the case of no pairs $P_{\rm jet}\sim L_{\rm d}$, and is a factor 4 smaller than $L_{\rm d}$ 
for the minimum number of protons allowed by Eq. \ref{ratiop}.
The ratio $n_\pm/n_{\rm p}$ behaves approximately as in Fig. \ref{lmin}.

\section{Conclusions}

It is very likely that the $\gamma$--ray emission region in powerful FSRQs is within
their broad line region, with broad line photons being the seeds for the inverse Compton
scattering process.
We base this assumption on the observed fast variability, difficult to explain
in models where dissipation takes place at much larger distances in the jet, as in the models
by Marscher et al. (2008) and Sikora et al. (2008; 2009).
This implies that the radiation is anisotropic in the comoving frame, making
the jet to recoil.
The observer would then see a deceleration of the jet, important for 
pure electron--positron light jets and becoming less significant if 
the jet is heavier due to the presence of protons.
Therefore the requirement of no or only modest deceleration
translates in a requirement on the amount of protons in the jet.
This then gives a lower limit on the total jet power.

Within the framework of synchrotron and external Compton models,
the only parameters that remain somewhat free (namely not accurately given by
observational data) for calculating the minimum $P_{\rm jet}$ are 
the viewing angle $\theta_{\rm v}$ and the bulk Lorentz factor $\Gamma$.
We can however see how the minimum $P_{\rm jet}$ values change as a function
of $\Gamma$, assuming a given viewing angle.
Doing so, we find at which $\Gamma$ the jet power is minimized.
We can then compare this minimum $P^{\rm min}_{\rm jet}$ (``minimum of the minimum values")
with the accretion disk luminosity.

Applying these arguments to 3C 454.3, one of the best studied $\gamma$--ray blazars,
we found that if $\theta_{\rm v}=1/\Gamma$, then $P^{\rm min}_{\rm jet}>L_{\rm d}$,
while it becomes a factor 4 smaller than $L_{\rm d}$ if $\theta_{\rm v}=0^\circ$.
The key question if the jet power can be larger than the accretion disk luminosity
remains therefore open, but with a narrower range of possibilities than before.
We can exclude pure electron--positron jets, but we can allow for $\sim10$
pairs per proton. 
This value is in agreement with what found by Sikora \& Madejski (2000)
using a different argument.
Be aware that all these estimates are based on the assumption that {\it all} leptons
present in the source participate to the emission.
If cold leptons were present, they would increase our estimate of the jet power.

\section*{Acknowledgements}
We thank G. Ghirlanda and L. Sironi for useful discussions.

\end{document}